\documentclass[preprint,preprintnumbers,amsmath,amssymb]{revtex4}

\usepackage{graphicx}
\usepackage{dcolumn}
\usepackage{bm}
\usepackage{color}

\makeatletter       

\renewcommand{\p@subsection}{}

\renewcommand{\p@subsubsection}{}

\makeatother

\begin{document}

\preprint{BA-05-104}

\title{Classical Cancellation of the Cosmological Constant Re-Considered}

\author{Stephen M. Barr}%
\email{smbarr@bartol.udel.edu}%
\affiliation{Bartol Research Institute, University of Delaware,
Newark, Delaware 19716, USA.}

\author{Siew-Phang Ng}%
\email{spng@bartol.udel.edu}%
\affiliation{Bartol Research Institute, University of Delaware,
Newark, Delaware 19716, USA.}

\author{Robert J. Scherrer}%
\email{robert.scherrer@vanderbilt.edu} \affiliation{Department of
Physics and Astronomy, Vanderbilt University, Nashville, TN 37235,
USA}

\date{\today}

\begin{abstract}
We revisit a scenario in which the cosmological constant is
cancelled by the potential energy of a slowly evolving scalar
field, or ``cosmon". The cosmon's evolution is tied to the
cosmological constant by a feedback mechanism. This feedback is
achieved by an unconventional coupling of the cosmon field to the
Ricci curvature scalar. The solutions show that the effective
cosmological constant evolves approximately as $t^{-2}$ and
remains always of the same order as the density of ordinary matter
and radiation. Newton's constant varies on cosmological time
scales, with $\dot{G}_N/G_N \ll 1/t$. $G_N$ could have been
somewhat different, and possibly smaller, at the time of Big Bang
nucleosynthesis.
\end{abstract}

\maketitle

\section{Introduction}

Many ideas have been suggested for solving the cosmological
constant problem \cite{review}, but none so far has proven to be
completely satisfactory. An old idea is that the vacuum energy of
ordinary fields is cancelled by some compensating field, sometimes
called a ``cosmon". The cosmon would roll down a potential hill
until it had lowered the total vacuum energy to an acceptable
level \cite{barr, ford, cosmon}. The question is how the cosmon
field would ``know" when to stop rolling. If the cosmon potential
happened to have a minimum at exactly the right height, it can
cancel off other contributions to $\Lambda$. However, that would
obviously be nothing other than a manual setting of the total
$\Lambda$ to zero at the minimum, i.e. just the old fine-tuning of
$\Lambda$ that one is trying to avoid.

If a conspiracy of parameters is to be avoided, it is imperative
that the cosmon action does not directly depend on the
contributions to $\Lambda$ coming from other sectors of the
theory. But how, in that case, is the cosmon to know when to stop
rolling? Some sort of feedback mechanism is required. Several such
mechanisms have been proposed over the years \cite{barr, ford,
cosmon, feedback}. The one we will explore in this paper was
proposed in the papers of Ref. \cite{barr,ford} in the mid 1980s,
and is based on the idea that the cosmon field couples to the
scalar curvature $R$. Since the trace of the Einstein field
equations tells us that $-R = 2 \Lambda + \frac{1}{2} (\rho - 3
p)$, the cosmon can learn about $\Lambda$ through its effect on
$R$. The conventional wisdom, based on a ``no-go theorem" stated
in Weinberg's well-known 1989 review of the cosmological constant
problem, is that this kind of feedback mechanism cannot work.
However, this ``theorem", which originally comes from the papers
\cite{barr,ford}, does not apply to the kind of model we discuss
in this paper, as we shall see. (Note: we will be using units
where $16 \pi G_N = 1$, signature -+++, and the sign convention in
which a sphere has negative $R$.)

Before exploring this idea further, there is a general point that
may be important. In any realistic feedback mechanism, it would
seem that the ``effective vacuum energy" $\lambda$ --- by which we
mean the vacuum energy from other sectors of the theory
($\Lambda$) plus the potential energy in the cosmon field
($V_{cosmon}$) --- must fall to zero as time passes in such a way
that it is always of roughly the same order as the matter energy
density, $\rho_{matter}$. By ``matter" here we mean all forms of
particles, including baryons, photons, neutrinos, dark matter
particles, etc. The reason for this is as follows. If $\lambda$
falls off significantly more {\it slowly} than $\rho_{matter}$,
then it will quickly come to dominate the universe, which will end
up with $\Omega_{matter}$ being negligibly small. If, on the other
hand, $\lambda$ falls off significantly {\it faster} than
$\rho_{matter}$, then very early in the history of the universe
the effective cosmological constant would be much smaller than
$T^4$, where $T$ is the temperature of the universe. However, that
would mean that any subsequent phase transition (such as one
associated with the symmetry breaking of grand unification, the
weak interactions, or the chiral symmetry of the quarks) would
drive $\lambda$ very negative, since it would lower $\Lambda$ by
an amount of order $T^4$. Given the arguments of Coleman and De
Luccia \cite{coleman} and more recently from a string theoretic
viewpoint, Banks \cite{banks}, it is questionable whether such a
transition could occur. But if it did, once the universe has
negative $\lambda$, it is not clear how it could get out of that
hole, as it would have to in order to be consistent with present
observations.

The upshot is that any successful feedback mechanism would seem to
require that the hidden energy density of $\Lambda$ plus ``cosmon"
field energy must be roughly of order the density of ordinary
matter at all times. This is very suggestive, given the puzzling
fact that a dark energy has been observed that is comparable to
the matter energy today.

The feedback model we discuss does indeed have the desired feature
that $\lambda \sim \rho_{matter}$ at all times. As we shall see,
however, this does not necessarily account for the dark energy
that has been observed. The problem is that any mechanism that is
designed to erase $\Lambda$ will tend to erase dark energy too. In
the model we discuss there is indeed energy that is ``dark"
(namely $\lambda = \Lambda + V_{cosmon}$), which generally falls
to zero roughly as $t^{-2}$ and has $w \cong 1/3$. However, after
a phase transition $\lambda$ tends to fall more slowly for a
period of time, so that $w$ can be smaller than $1/3$ and perhaps
even close to $-1$. It is possible that this fact can be used to
account for the nearly constant dark energy that has recently been
observed.

To return to the specific idea that the feedback occurs through
coupling of the cosmon to the scalar curvature $R$, the obvious
way for this to happen is through the cosmon potential. To
illustrate how this might work, consider a toy model
\cite{barr,ford} with $V_{cosmon} = - \alpha \phi R$, where $\phi$
always denotes the cosmon in this paper. The effective vacuum
energy, as we defined it above, is then $\lambda = \Lambda -
\alpha \phi R$. As the cosmon rolls down its linear potential
hill, $\lambda$ will decrease; and as that happens, the scalar
curvature will decrease as well. This decrease of the scalar
curvature reduces the slope ($- \alpha R$) of the cosmon
potential, and therefore the cosmon rolling will decelerate (since
there is friction due to the Hubble expansion). If we assume that
$R$ receives contributions only from $\lambda$, then as $\lambda$
approaches zero, the slope of the cosmon potential will approach
zero also, and the velocity of the cosmon's rolling will approach
zero. In other words, whatever the value of $\Lambda$ is, the
quantity $\lambda$ will asymptotically approach zero as time
passes. That is, the potential energy of the cosmon, $- \alpha
\phi R$ will progressively erase $\Lambda$.

In fact, this is what actually would happen in this toy model, as
an explicit solution of the field equations shows. Unfortunately,
however, Newton's constant $G_N$ suffers a disaster. Instead of
the usual Einstein-Hilbert action $-R$, one has $-(1 - \alpha
\phi) R$. Since the quantity $\alpha \phi R_U$ (where $R_U$ is the
background scalar curvature of the universe) is asymptotically
approaching $\Lambda$, that means that $(16 \pi G_N)_{eff}^{-1} =
1 - \alpha \phi$ is asymptotically approaching $1 - \Lambda/R_U$,
which is exponentially large at present ($\sim 10^{120}$). In
other words, Newton's ``constant" turns off in a drastic way.

This is the problem that seems to afflict any attempt to construct
a feedback mechanism through a coupling of the cosmon to the
scalar curvature through the cosmon potential. (This is the
conclusion that was reached in the papers \cite{barr,ford} and
repeated in Weinberg's 1989 review \cite{review}.) This led to an
attempt in \cite{barr} to achieve a viable feedback through the
{\it kinetic} term of the cosmon. What is required is that
negative powers of $R$ appear in the cosmon kinetic term, which,
of course, is a peculiar thing indeed \cite{overR}. Nevertheless,
a model that erases the cosmological constant without any
fine-tuning of parameters at the classical level can easily be
constructed as shown in \cite{barr}. In the paper we correct and
extend the analysis in that earlier paper, and address several
important issues that were not considered there.

This paper is organized as follows. In section 2 we present the
toy model that we are discussing and find cosmological solutions
for the cases where cosmon energy is dominant and where radiation
is dominant. (In \cite{barr} only the cosmon-dominated case was
considered, and the analysis of it was faulty, since certain
assumptions were made that were not self-consistent.) We also
discuss the stability of the solutions, and what the cosmon
equation of state looks like. We find that while there are
modifications to the effective Newton's constant (because of the
appearance of $R$ in the cosmon kinetic term), they are of order 1
and harmless.

In section 3, we look at what happens when baryonic matter (or
other matter whose stress-energy tensor is not traceless) is
present. We confirm the conclusion of \cite{barr} that this does
not disturb the feedback mechanism for erasing the cosmological
constant. However, there is an interesting effect: the cosmon
field does wipe out the trace of the stress-energy tensor inside
baryonic matter (contrary to what was claimed in \cite{barr}).
That is, inside a lump of baryonic matter (e.g. a nucleon, atomic
nucleus, or neutron star) the cosmon field has a pressure and
density such that the total stress-energy tensor is traceless.
However, this has less dramatic effects than one might suppose.
This will be analyzed in section 3.

In section 4, we discuss other issues, such as stability of
solutions, what happens if the vacuum energy is suddenly reset by
a phase transition, and quantum corrections to the model and
fine-tuning.

In the Appendix we point out certain mistakes and deficiencies in
the analysis in \cite{barr}.

\section{The Toy Model}

The model that we shall discuss in the rest of this paper has the
action
\begin{equation}
{\cal L} = - \frac{c^2}{2} (\partial_{\sigma} \phi
\partial^{\sigma} \phi)R^{-2} + \alpha \phi - R - \Lambda_0 +
{\cal L}_{matter},
\end{equation}

\noindent where $c$ is a parameter with dimensions of $M^4$ that
we take to be of order one in Planckian units. $\Lambda_0$ is the
vacuum energy that we are trying to erase. (It may change suddenly
if there is a first order phase transition.) The third term is
just the usual Einstein-Hilbert action. Our metric signature is
$(-+++)$. We use the convention that $R$ is negative for a sphere.

The idea is the following. As the cosmon rolls down its potential
hill, the effective vacuum energy $\lambda = \Lambda_0 - \alpha
\phi$ will decrease, causing $-R$ to decrease also. This in turn
will cause the kinetic term of the cosmon to stiffen, so that the
rolling of the cosmon will slow down. As $R$ approaches zero, so
does the velocity of the cosmon. In fact, even though $R$
approaches zero, $\dot{\phi}$ approaches zero faster, so that the
kinetic energy of $\phi$ becomes smaller, not larger, even though
$R$ is in the denominator. Our explicit solutions will show this.

Having a negative power of $R$ in the cosmon kinetic term
\cite{overR} can be avoided if we use an auxiliary field:
\begin{equation}
\begin{array}{cl}
{\cal L} = & - \frac{1}{2} X^2 (\partial_{\sigma} \phi
\partial^{\sigma} \phi)
+ \alpha \phi \\ & \\
& +  (\partial_{\sigma} X \partial^{\sigma} X)/X^m - bR(XR - c)^2 \\ &  \\
& -   R - \Lambda_0 + {\cal L}_{matter}.
\end{array}
\end{equation}

\noindent The power $m$ in the kinetic term of the auxiliary field
$X$ must be large enough so that $X$ can adjust rapidly to changes
in $R$. For large $b$ and $m$ this version of the model reduces to
that in Eq. (1). We shall therefore use Eq. (1).

The gravity equation coming from Eq. (1) is
\begin{equation}
\begin{array}{ccl}
G^{\lambda \rho}(1 - f) & = & -\frac{1}{2} T^{\lambda
\rho}_{matter}
+ \frac{1}{2} \lambda g^{\lambda \rho} \\ & & \\
& - & \frac{c^2}{2} (\partial^{\lambda} \phi \partial^{\rho} \phi
- \frac{3}{2} \partial^{\sigma} \phi \partial_{\sigma} \phi
g^{\lambda \rho})
R^{-2} \\ & & \\
& + & (g^{\lambda \rho} \Box - \nabla^{(\lambda} \nabla^{\rho)})
f,
\end{array}
\end{equation}

\noindent where
\begin{equation}
f \equiv c^2 \partial^{\sigma} \phi \partial_{\sigma} \phi R^{-3},
\end{equation}
\noindent and
\begin{equation}
\lambda \equiv \Lambda_0 - \alpha \phi,
\end{equation}
\noindent which implies, of course, that $\dot{\lambda} = - \alpha
\dot{\phi}$. The last term in Eq. (3) arises from $\frac{\delta
R}{\delta g_{\lambda \rho}} \frac{\delta}{\delta R} \left( -
\frac{c^2}{2} \partial_{\sigma} \phi \partial^{\sigma} \phi R^{-2}
\right)  = (\delta R/\delta g_{\lambda \rho}) f$ after two
integrations by parts. As we shall see, when there is only cosmon
energy, the combination of fields that we have called $f$ ends up
being simply a constant, so that the last term in Eq. (3)
vanishes. When there is matter present, however, $f$ is not
constant and the last term in Eq. (3) becomes very important, as
we shall see.

If we take the trace of Eq. (3), we get
\begin{equation}
\begin{array}{l}
- R (1 + \frac{3}{2} f) = -\frac{1}{2} T_{matter} + 2 \lambda + 3
\Box f.
\end{array}
\end{equation}
\noindent The equation of motion for $\phi$ is
\begin{equation}
- \frac{1}{\sqrt{-g}} \partial_{\sigma}\left[ \sqrt{-g} (c^2
\partial^{\sigma} \phi R^{-2}) \right] = \alpha, \label{ScalarEOM}
\end{equation}
\noindent and assuming a flat Friedmann-Robertson Walker (FRW)
ansatz for the metric, the expression for the scalar curvature in
terms of the scale factor of the universe $a(t)$ is, with our sign
conventions,
\begin{equation}
-R = 6\left( \frac{ \ddot{a}}{a} + \left( \frac{\dot{a}}{a}
\right)^2 \right).
\end{equation}

We will solve these equations first in the case where there is no
ordinary matter, but only cosmon energy. Let us assume that $a(t)
\propto t^n$  and that the fields are constant in space. Then the
cosmon equation of motion becomes $t^{-3n}
\partial/\partial t \left[ t^{3n} \dot{\lambda} R^{-2} \right] = -
\alpha^2/c^2$, and has the solution
\begin{equation}
\begin{array}{l}
\dot{\lambda} R^{-2} = - \frac{1}{3n+1} (\alpha^2/c^2) t - k
t^{-3n}.
\end{array}
\end{equation}
\noindent The second term on the right is a transient that dies
off rapidly. (It has already implicitly been assumed to be
negligible, since its presence would cause the scale factor to
deviate from a pure power dependence on $t$.) It is easily shown
that the solution to the coupled equations gives $R$ proportional
to $\lambda$, so we will write $-R = 2 \gamma \lambda$, with the
number $\gamma$ to be determined. Substituting this into Eq. (9)
and dropping the transient term gives $\lambda = \frac{3n+1}{2
\gamma^2} (c^2/\alpha^2) t^{-2}$, for large $t$ \cite{t2}. From
this, it is easy to see that $f$ is independent of $t$, so that
the last term in Eq. (6) may be dropped. Since we are assuming
that there is no matter present, that equation then takes the
simple form $-R(1 + \frac{3}{2} f) = 2 \lambda$, which implies
that $(1 + \frac{3}{2} f) = \gamma^{-1}$. We can get another
relation between $f$ and $\gamma$ from the definition of $f$, Eq.
(4), which we can write as $f = (2 \gamma)^{-3} (c^2/\alpha^2)
\dot{\lambda}^2 \lambda^{-3}$. Substituting our solution for
$\lambda$ into this gives $f = (3n+1)^{-1} \gamma^{-1}$. Combining
the two expressions relating $f$ and $\gamma$ gives $\gamma =
\frac{6n-1}{2(3n+1)}$ and $f = \frac{2}{6n-1}$. Therefore
\begin{equation}
\begin{array}{l}
\lambda = \frac{2 (3n+1)^3}{(6n-1)^2} \frac{c^2}{\alpha^2} t^{-2}.
\end{array}
\end{equation}

It remains to use Eq. (8), which tells us that $-R = 12
n(n-\frac{1}{2}) t^{-2}$. Our solution for $\lambda$ and $\gamma$
gives $-R = 2 \gamma \lambda = \frac{2(3n+1)^2}{6n-1}
\frac{c^2}{\alpha^2} t^{-2}$. Combining these gives the power $n$
in terms of the parameters of the model:
\begin{equation}
\frac{6n(6n-1)(n-\frac{1}{2})}{(3n+1)^2} = \frac{c^2}{\alpha^2}.
\end{equation}
Note that physical quantities must depend on the combination
$c^2/\alpha^2$ since it is invariant under rescalings of the
fields. Let us call this combination $\epsilon$. If we take
$\epsilon \ll 1$, then
\begin{equation}
\begin{array}{l}
n \cong \frac{1}{2} + \frac{25}{24} \epsilon.
\end{array}
\end{equation}
\noindent Note that the universe expands faster than $t^{1/2}$. If
there is a small amount of radiation in this cosmon-dominated
universe, it will redshift as $\rho_{rad} \sim t^{-2 -\frac{25}{6}
\epsilon}$, whereas the ``effective vacuum energy", $\lambda$,
goes as $t^{-2}$. We shall find that the same is true in a
radiation dominated universe with a small amount of cosmon energy:
the cosmon/radiation ratio grows as a power of $t$, where the
power is of order $\epsilon$. Thus if $\epsilon \sim 10^{-2}$,
say, the cosmon energy tracks the radiation density very well over
many orders of magnitude in time.

The quantity $\lambda$ contains the vacuum energy $\Lambda_0$
(i.e. the underlying cosmological constant) and the cosmon
potential energy. If one considers the full ``dark" stress energy,
which includes the contributions from $\Lambda_0$ and from both
the kinetic and potential energy of the cosmon, one finds that for
$\epsilon \ll 1$, $p \cong \rho/3$. In other words, all the
``dark" energy --- cosmological constant plus cosmon --- has an
equation of state similar to radiation.

What can we say about Newton's constant? The coefficient of the
Einstein tensor in Eq. (3), which we may call $(16 \pi
G_N)_{eff}^{-1}$, is given by $(1-f) = 6(n- \frac{1}{2})/(6n-1)$,
which for small $\epsilon$ is approximately $\frac{25}{8}
\epsilon$. What matters, of course, is not the particular value of
$(G_N)_{eff}$, which is just a meaningless rescaling of the Planck
scale, but how $(G_N)_{eff}$ changes with time. We see that if
only cosmon energy (including $\Lambda$) is present $(G_N)_{eff}$
is constant. We shall see shortly that when matter is present,
$(G_N)_{eff}$ changes over cosmological time scales, but very
little if $\epsilon \ll 1$, which is the case of interest.

Now let us turn to the case of a radiation-dominated universe,
with some cosmon energy. Since radiation is dominant, one expects
the scale factor to grow approximately as $t^{1/2}$, so we take
\begin{equation}
a(t) = a_0 (t/t_0)^{1/2} (1 + g_1 (t/t_0)^d + g_2 (t/t_0)^{2d} +
...),
\end{equation}
\noindent where $t_0$ is some time that, as we shall see, is close
to when the cosmon energy is equal to the radiation energy. The
power $d$ is negative, as is $g_1$, so that the cosmon/radiation
ratio decreases. However, since $d$ turns out to be of
$O(\epsilon)$, the cosmon/radiation ratio will change very slowly
for small $\epsilon$. Eq. (13) gives
\begin{equation}
-R = 6 d(1+d) g_1 t_0^{-2} (t/t_0)^{-2+d} + O(t^{-2+2d}).
\end{equation}

\noindent It is easy to show that the leading behavior of
$\lambda$ is $t^{-2 +2d}$, so we write
\begin{equation}
\lambda = \lambda_0 t_0^{-2} (t/t_0)^{-2+2d} + O(t^{-2 +3d}).
\end{equation}
\noindent Substituting Eqs. (14) and (15) into the cosmon equation
of motion, Eq. (7), we find $\lambda_0 = \frac{36}{5}
\frac{\alpha^2}{c^2} \frac{d^2 (1 +d)^2}{1-d} g_1^2$. Substituting
Eqs. (14) and (15) into the Eq. (4) (the definition of $f$), we
find $f = \frac{24}{25} d(1+d) (\alpha^2/c^2) g_1 (t/t_0)^d$ to
leading order.

Considering now the trace of the gravity equation, Eq. (6), we see
that the leading terms on both sides go as $t^{-2 +d}$. If we
match only the leading terms (which is good enough for our
purposes), we may neglect $\lambda$, which contains only terms of
order $t^{-2 +2d}$ and higher.  We may also neglect the
$\frac{3}{2} f$ on the left-hand side for the same reason. Since
the matter is assumed to be radiation, $T_{matter} = 0$. With
fields assumed to be constant in space, we need keep only time
derivatives, so that Eq. (6) reduces finally to the simple form
$-R = -3 (\frac{\partial^2}{\partial t^2} + \frac{3}{2t}
\frac{\partial}{\partial t}) f = -3 d (d+\frac{1}{2}) f t^{-2}$,
which gives $-R = -\frac{72}{25} d^2 (d+ \frac{1}{2})(1 + d)
(\alpha^2/c^2) g_1 t_0^{-2} (t/t_0)^{-2+d}$. Combining this with
Eq. (14) yields
\begin{equation}
\begin{array}{l}
-\frac{6}{25} d (1+2d) = \frac{c^2}{\alpha^2} = \epsilon.
\end{array}
\end{equation}
For small $\epsilon$, $d \cong -\frac{25}{6} \epsilon$, so that
\begin{equation}
\lambda/\rho_{radiation} \sim t^{-\frac{25}{3} \epsilon}.
\end{equation}
\noindent In fact, since the leading term in Einstein's equations
gives $\rho_{radiation} \cong \frac{3}{2} t^{-2}$, one has
$\lambda/\rho_{radiation} \cong \lambda_0 (t/t_0)^{2d}$. The
preceding calculations give the result $\lambda_0 = -\frac{30 d (1
+ d)^2}{(1-d)(1+2d)} g_1^2$. Consequently, for small $\epsilon$,
one has $\lambda/\rho_{radiation} \cong \frac{250}{3} \epsilon
g_1^2 (t/t_0)^{2d}$. We see, then, that when $\lambda$ is
comparable to the radiation density, the terms in $a(t)$ of higher
order in $t^d$ are of the same size as the lower order terms, and
the whole approximation scheme we have been using breaks down, as
one would indeed have expected.

Note that the effective Newton's Constant in an era where
radiation dominates is (for small $\epsilon$) given by $(16 \pi
G_N)_{eff}^{-1} = 1-f \cong 1 + 4 g_1 (t/t_0)^{-\frac{25}{6}
\epsilon}$. That is, $G_N$ decreases very slowly, with
$\dot{G_N}/G_N \sim O(\epsilon)/t$ (although see the discussion in
Sec. 4 below).

\section{What Happens Inside Matter}

The mechanism we have been describing relies on the fact that the
scalar curvature is a ``proxy" for the vacuum energy. However,
inside a matter distribution there is also a contribution from the
trace of the matter stress-energy, as shown in Eq. (6). One might
worry that this would interfere with the mechanism for cancelling
the cosmological constant.

To study this situation, let us consider a spherical mass of
radius $r_m$, and constant density $\rho_m$, surrounded by a
vacuum. We take $\rho_m$ to be nuclear density, for specificity,
and the pressure to be negligible. This mass could be a baryon, or
a neutron star, for example. One might suppose that since in
natural gravitational units, $\rho_m \sim 10^{-80}$ while $\lambda
\sim 10^{-120}$, that $\rho_m$ would dominate the right-hand side
of Eq. (6). Let us make this assumption to start with, even though
it will prove to be false, because it will be instructive and
point us toward the right solution.

Using spherical symmetry, we can write the equation of motion for
the cosmon inside the mass as $\frac{1}{r^2}
\frac{\partial}{\partial r} ( r^2 \frac{\partial \lambda}{\partial
r} ) = \frac{1}{4} \rho_m^2 (\alpha^2/ c^2)$. This has the
solution, $\lambda = \lambda_{vac} + \frac{1}{24} \rho_m^2
(\alpha^2/c^2) (r^2 - r_m^2)$, for $r \leq r_m$, where
$\lambda_{vac}$ is the value of $\lambda$ in the vacuum, which we
have given in Eq. (10). Then $\Delta \lambda \equiv \lambda(r=0) -
\lambda_{vac} \sim \rho_m^2 r_m^2$. If $r_m$ is the radius of a
baryon, this is of order $10^{-120}$ in natural units. Even if
$r_m$ is the radius of a neutron star, $\Delta \lambda$ is only of
order $10^{-80}$. Now consider the quantity $f$. Inside the sphere
of matter, $f \cong \frac{c^2}{\alpha^2} (\frac{\partial
\lambda}{\partial r} )^2 R^{-2} \sim \rho_m^2 r_m^2$, which for a
baryon is $O(10^{-120})$ and for a neutron star is $O(10^{-80})$.
If our assumptions were correct, then, it would imply that $f
\cong 1$ far from the matter and $f \ll 1$ inside. But this cannot
be the case, for then somewhere inside or near the matter
distribution the quantity $\nabla^2 f$ would have to be of order
$1/r_m^2$. For a baryon this is $O(10^{-40})$ which is many orders
of magnitude larger than nuclear density. That is, the last term
in Eq. (6) would greatly dominate, contrary to our assumption.
Thus the assumption that $\rho_m$ dominates is false.

This shows what the real solution of the coupled equations must
look like. It must be that $f = f_{vac} + \tilde{f}(x^{\mu})$,
where $\tilde{f} \ll f_{vac} \ll 1$ and $\nabla^2 \tilde{f} \cong
- \frac{1}{6} \rho_m$ (inside matter) so that the terms
$-\frac{1}{2} T_{matter}$ and $3 \Box f$ on the right-hand side of
Eq. (6) cancel. Then $-R$ can have the same value inside and
outside the matter distribution, allowing $\lambda$ to be nearly
spatially constant. Consequently, $f \cong f_{vac}$ even inside
matter. This is the only way to get a self-consistent solution.

The net result is the following. The effective Newton's constant
$(16 \pi G_N)_{eff}^{-1} = 1-f$ is the same everywhere, both
inside baryonic matter (or any other kind of matter) and in the
vacuum. This is what we would expect anyway from the Equivalence
Principle. However, something very strange happens; namely the
total stress energy tensor acts like radiation, since the sum of
the matter stress energy and the last term in Eq. (3), namely
$(g^{\lambda \rho} \Box  - \nabla^{(\lambda} \nabla^{\rho)}) f$,
gives something traceless. It is easy to see that if the mass of a
ball of baryonic matter is $M$ due to $T^{\lambda \rho}_{matter}$,
the total mass including the last term in Eq. (6) is only
$\frac{2}{3} M$.

Isn't it catastrophic to have the total stress energy be
traceless, even for ordinary matter? For example, consider a box
of gas. The spatial average of the pressure inside the box would
be $\frac{1}{3} \rho_{gas}$, where $\rho_{gas}$ includes the rest
energy of the gas molecules. However, it must be realized that
this average pressure is almost all coming from pressure in the
cosmon field {\it inside the material particles}. The actual gas
pressure due to particles colliding with each other is just what
would be given by the kinetic theory of gases. On the other hand,
the box of gas would gravitate as though it had traceless stress
energy.

\section{Other Issues}

The cosmological solutions we have found are stable under small
perturbations. This can be seen from the solution to the cosmon
equation of motion given in Eqs. (9) and (10). The first integral
of the equation of motion depends on an integration constant
denoted $k$ in Eq. (9). One sees that $k$ describes a transient
term that dies away with increasing $t$. The integration of Eq.
(9) gives another integration constant that we have not shown in
Eq. (10). However, the exact solution approaches the one given in
Eq. (10) for large $t$.

There is a simple physical reason why the solutions we have given
are stable under small perturbations. Imagine that $\lambda$ finds
itself at some $t$ with a somewhat smaller value than that given
in Eq. (10). Then $R$ will also be somewhat smaller than in our
solutions. That will cause the kinetic term of the cosmon to be
more ``stiff", which causes the cosmon field to roll down its
potential hill more slowly than it does in our solution. In other
words, if $\lambda$ is initially "too small", it compensates by
decreasing more slowly, until eventually it rejoins the asymptotic
solution given in Eq. (10). In the same way if $\lambda$ starts
off somewhat larger than the solution in Eq.(10), then $R$ is
larger and the cosmon rolls down its hill faster until it rejoins
the asymptotic solution.

\begin{figure}
\includegraphics[scale=0.9]{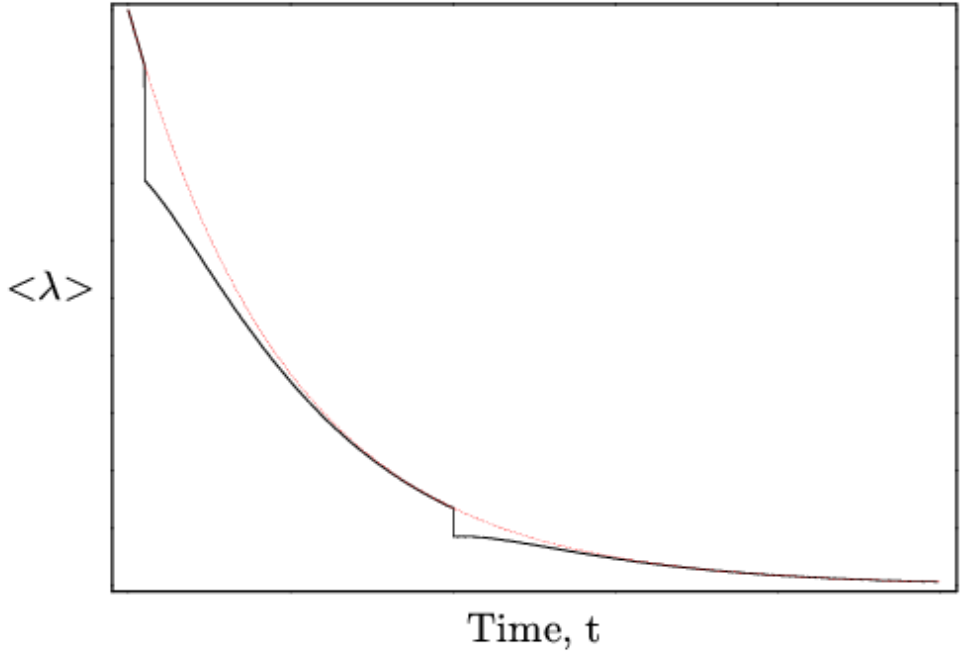}
\caption{\label{FigSchema}Schematic diagram of the evolution of
the cosmon field in the presence of phase transitions. The solid
line represents the actual evolution of the cosmon field while the
dashed line is the trajectory it would assume in the absence of
any phase transitions. In the above case, we have included only
two phase transitions; after both of which, the vev of the cosmon
field asymptotes back to the pre-phase transition trajectory.}
\end{figure}

This feature has an important consequence that would be helpful in
making a realistic model of cosmology. In a realistic particle
theory there would be phase transitions that change the vacuum
energy density relatively suddenly, such as those connected with
the breaking of grand unified symmetries, supersymmetry, the weak
interaction symmetries, or the chiral symmetries of the quarks.
Thus, the cosmological ``constant" $\Lambda_0$ that has to be
cancelled off by any cosmon field undergoes sudden jumps downward.
The stability of our solutions allows our mechanism to accomodate
these jumps. Suppose that $\lambda \equiv \Lambda_0 - \alpha \phi$
is evolving as shown in Eq. (10), namely as $t^{-2}$, and suddenly
a phase transition resets $\Lambda_0$ to be smaller by an amount
of order $T^4$, where $T$ is the temperature at which the phase
transition happens. Now $\lambda$ is of order $\rho_{radiation}$,
as we have seen; and $\rho_{radiation}$ is of order $T^4$. So
$\lambda$ will be reset downward by an amount that is of order its
own size. Let us suppose that it is not driven to a negative value
by this, but only to a smaller positive value. Then, for the
reasons we have just explained, the rolling of the cosmon down its
hill will be slowed until $\lambda$ rejoins the asymptotic
solution we have found. In fact, this could happen repeatedly at
various epochs in the history of the universe: the effective
vacuum energy could be repeatedly reset by phase transitions, and
then after a few e-foldings in time, rejoin its asymptotic
solution. Fig.\ref{FigSchema} schematically illustrates the
evolution of just such a cosmon field that undergoes several phase
transitions. The fact that $\lambda$ levels off for a period of
time after a phase transition and acts more like a cosmological
{\it constant} may be a possible explanation of the observed dark
energy if the universe has recently undergone a phase transition.
There is a complication that such an explanation would face,
namely that the radiation released in the phase transition would
tend to dominate over $\lambda$ for a period of time. How
$\lambda$ and $w$ would behave after a phase transition is
something that has to be more carefully studied numerically before
one can tell if this is a viable explanation of dark energy.

Another effect of the phase transition would be a transient time
dependence of $G_N$. Suppose that $\Lambda_0$ and therefore $R$ is
decreased suddenly (compared to t), then by the equation of
motion, Eq. \ref{ScalarEOM}, one expects $\dot{\phi} R^{-2}$ to be
approximately constant during this sudden decrease in R. Thus,
$f=-c^2 \dot{\phi} ^2 R^{-3}$ is driven to a smaller value than
the asymptotic solution. As $\lambda$ evolves to a rejoin the
asymptotic solution, $f$ will be increasing and hence so will
$G_N$. In other words, if we are in an era where $\lambda$ is
``recovering" from a phase transition (which might explain the
near constancy of $\lambda$), then $G_N$ might be slowly
increasing. This can have implications for Big Bang
Nucleosynthesis.

 It should be noted that if a phase transition were to
drive $\lambda$ negative, there would be a catastrophe: $\lambda$
could not ever become positive again. Rather it would roll to ever
more negative values, which would make $|R|$ larger and the cosmon
kinetic term {\it less} stiff, so that the rolling would
accelerate and $\lambda$ would plunge ever faster into the abyss.
However, there does not seem to be any reason why phase
transitions would necessarily have to drive $\lambda$ negative.
Whether or not one did so would depend on the parameters of the
theory.

There are of course many other issues that would have to be
examined to see if a realistic cosmology could emerge from the toy
model we have described or some similar model. We intend to return
to these in the future. Another very important issue is whether
the kind of theory we are describing could make sense at the
quantum level. One obvious issue is whether there is any
fine-tuning is involved in the choice of cosmon kinetic term we
have made. If one considers the effects of either matter loops or
cosmon loops, these do not seem to raise any problems of fine
tuning (though we have not proved this). One thing that helps in
this regard is that the cosmon only couples to gravity. However,
simple power-counting arguments suggest that graviton loops induce
a term of the form $(\partial \phi)^2/R^3$ in addition to
$(\partial \phi)^2/R^2$. The former term would overwhelm the
latter unless its coefficient were tuned to be less than about
$10^{-120}$, i.e. about the same tuning that is involved in the
cosmological constant problem itself. However, even if this is the
case, that does not mean that the model achieves nothing. It seems
to be a solution to the cosmological constant problem at the {\it
classical} level (whether a realistic one or not, that remains to
be seen). That is, it can solve the ``small number problem" even
if it does not solve the ``fine tuning problem". However, any
conclusions about the quantum theory are premature, as we have not
studied it in detail.

\section{Conclusions}

We have analyzed a toy model that realizes the idea of a feedback
mechanism for cancelling the cosmological constant at the
classical level. The model has several appealing features. First,
the total ``dark energy" of vacuum energy plus cosmon field energy
is always of the same order as the energy in radiation and matter.
This is certainly suggestive, given the observed ``cosmic
coincidence" that the dark energy is currently of the same order
as the energy in particles. Second, the mechanism works even if
there are several phase transitions throughout the history of the
universe that reset the cosmological constant. As explained in the
introduction, any successful cancellation mechanism for the
cosmological constant must have these features. There are also two
interesting modifications of gravity. (a) The effective Newton's
constant changes slightly over cosmological time scales during
periods when cosmon energy is subdominant. (b) The cosmon stress
energy acts to make the total stress energy tensor approximately
traceless everywhere, even inside baryonic matter. There are
several aspects of the model that are less attractive. The most
serious is that the scalar curvature appears (at least
effectively) in the denominator of the cosmon kinetic term. This
makes it questionable whether this model would make sense at the
quantum level. Moreover, naive power counting suggests that the
model must be fine tuned at the one loop level by the same amount
that is involved in solving the cosmological constant ``by hand".
There is also the difficulty that the rolling of the cosmon field
tends to erase any dark energy that has $w \cong -1$ just as it
erases a true cosmological constant. In other words, the
cancellation mechanism works {it too} well! However, after phase
transitions, the cosmon field tends to stop rolling for a while,
so that the ``effective cosmological constant" is approximately
constant and may explain the observed dark energy if the universe
has recently undergone a phase transition. This deserves further
study.

There are several other aspects of the model that also require
further study. It must be seen what the implications of the
tracelessness of the stress-energy tensor are. The cosmology of
the model must be studied to see how close to being realistic it
can be made. And whether the model makes sense as a quantum theory
needs to be investigated. Finally, it would be interesting to see
if the feedback idea can be made to work in some other way that
does not involve such an exotic cosmon kinetic term.

\newpage

\begin{acknowledgments}
We would like to thank Alberto Iglesias and Takemichi Okui for
useful discussions. S.M.B. and S.P.N. were supported by the
Department of Energy under contract DE-FG02-91ER40626 while R.J.S.
was supported in part by the Department of Energy under contract
DE-FG05-85ER40226.

\end{acknowledgments}

\appendix
\section*{Appendix}

The present paper is a reconsideration of ideas first presented in
a 1987 paper of Barr \cite{barr}. The purpose of this appendix is
to point out a number of flaws in the analysis in that paper.
Before discussing them, one should note that in most of that
paper, an auxiliary field $X$ was used as in Eq. (2) here. That
is, instead of the cosmon kinetic term depending on negative
powers of the scalar curvature $R$, it depended on positive powers
of $X$, where some potential $V(X,R)$ forced $X$ to be of order
$1/R$. The cosmon kinetic term assumed there was of the form
$\frac{1}{2} (\partial_{\lambda} \phi)^2 X^{\eta}$, where $\eta$
was allowed to be a free parameter for much of the analysis,
whereas in this paper we always assume that it is 2 (in fact, it
can be shown that it must be 2 for the power-law solution to
dominate).

(1) The most serious mistake in the analysis in \cite{barr} was to
assume that the scale factor of the universe went as $a(t) \sim
t^{1/2}$. In general, as we see from Eqs. (12) and (13) here, this
is not the case, except in the (unrealizable) limit $c^2/\alpha^2
\rightarrow 0$. The false assumption that $a \sim t^{1/2}$
together with the assumption that $V(X,R)$ is of the form $f(XR)$,
led to the result (in Eq. (A15) of \cite{barr}) that $(16 \pi
G_N)^{-1}_{eff} = \frac{3 (\eta -2)}{2 (2 \eta -3)}$, and (in Eq.
(A10) of \cite{barr}) that the effective cosmological constant
$\lambda \sim t^{-2/(\eta-1)}$. So, in order to have positive
Newton's constant, it had to be assumed in \cite{barr} that $\eta$
is larger than 2, implying that $\lambda$ falls off more slowly
than the radiation density (which under the assumption that $a
\sim t^{1/2}$ obviously goes as $t^{-2}$). Indeed, if $\eta = 2 +
\delta$, with $\delta \ll 1$, then $(16 \pi G_N)^{-1}_{eff} =
O(\delta)$ and $\lambda/\rho_{radiation} \sim t^{O(\delta)}$.

Now, even though the analysis of \cite{barr} was based on a wrong
assumption about $a(t)$, and the conclusions therefore are
invalid, some of those conclusions resemble the results of the
correct analysis presented here. In the present paper, we also
find that $\lambda$ falls off slower than the radiation density.
And if our parameter $\epsilon \equiv c^2/\alpha^2 \ll 1$, then
$(16 \pi G_N)^{-1}_{eff} = O(\epsilon)$ and $\lambda/
\rho_{radiation} \sim t^{O(\epsilon)}$.

In \cite{barr}, it was thought that to have positive Newton's
constant in the case $\eta =2$, a rather complicated form of
$V(X,R)$ had to be used (see section III of that paper). However,
that is not the case, as we have found here.

(2) Another mistake in the analysis of Appendix A of \cite{barr}
was in assuming that $V(X,R)$ could have the form $f(XR)$ (cf. Eq.
(A2) of \cite{barr}). This leads to an inconsistency in the
analysis as one sees from Eq. (A11) of that paper, which is
$\partial V/\partial X = \frac{2 \gamma}{\beta} \frac{2
\eta}{5(\eta-1)} \lambda^2$. The left-hand side is $f'(XR) R =
O(R)$, whereas the right-hand side is $O(\lambda^2) = O(R^2)$.
That is why in Eq. (2) of the present paper we chose the form
$V(X,R) = b R (XR-c)^2$, which is of the form $Rf(XR)$ and does
not lead to this inconsistency.

(3) There is a mistake in Eq. (A3) of \cite{barr}, in the last
term, which contains the operator $2 (g^{\lambda \rho} \Box -
\nabla^{\lambda} \nabla^{\rho})$. The factor of 2 should not be
present, and the indices on $\nabla^{\lambda} \nabla^{\rho}$
should be symmetrized. In other words, the operator should be the
same as in Eq. (3) of this paper.

(4) The analysis of what happens inside matter was incorrect in
that it assumed that the matter density dominates the right-hand
side of what is Eq. (6). As a result, it was thought that what we
call $f$ here (which corresponds to $\partial V/\partial R$ in
\cite{barr} if one makes use of the equation of motion for $X$)
varies from being very small inside matter to being of order one
outside. As was noted in \cite{barr}, this would make the terms
with second derivatives of $f$ (or equivalently $\partial
V/\partial R$) enormously large, so large as to dominate over the
matter density. As we found here, the resolution of this apparent
inconsistency is that $f$ inside matter is almost exactly the same
as outside, but its second derivatives almost exactly cancel the
trace of the stress energy tensor of the matter. This was not
realized in \cite{barr}, so that some of the discussion in section
IV of that paper is incorrect.

In spite of these flaws, most of the major conclusions of that
paper stand as correct, including the following: (i) in the model
where $R$ is contained in the cosmon potential Newton's constant
is driven to zero; (ii) this problem does not arise if $R$ couples
to the cosmon through the denominator of its kinetic term; (iii)
$\lambda$ falls off slower than the radiation or matter density if
$\lambda$ dominates; (iv) the feedback mechanism works even when
lumps of matter are present because the cosmon field does not
significantly ``sag" inside matter, and is therefore everywhere
approximately equal to its value in empty space. Our present
analysis shows that some apparent problems discussed in
\cite{barr} are not real. Most importantly, the disastrously large
contributions to the stress energy coming from the second
derivatives of $f$ (or $\partial V/\partial R$) do not arise, and
there is no difficulty in obtaining a positive Newton's constant.

\end{document}